\documentclass[
twocolumn,
superscriptaddress,
showpacs,preprintnumbers,
 amsmath,amssymb,
 aps,
prapplied,
longbibliography
]{revtex4-2}

\usepackage{graphicx}% Include figure files
\usepackage{dcolumn}% Align table columns on decimal point
\usepackage{bm}% bold math
\usepackage[update,prepend]{epstopdf}
\usepackage{xcolor}
\usepackage{ulem}
\usepackage{hyperref}% add hypertext capabilities
\usepackage{float}

\DeclareMathOperator{\sech}{sech}

\begin{document}

%\preprint{Arxiv version}

\title{Mode-locked waveguide polariton laser}

\author{H.~Souissi}
\affiliation{Laboratoire Charles Coulomb (L2C), Universit\'e de Montpellier, CNRS, Montpellier, France}

 \author{M.~Gromovyi}
 \affiliation{Centre de Nanosciences et de Nanotechnologies, CNRS, Universit\'e Paris-Saclay, France}
 \affiliation{UCA, CRHEA-CNRS, Rue Bernard Gregory, 06560 Valbonne, France}
 
 \author{I.~Septembre}
\affiliation{Universit\'e Clermont Auvergne, CNRS, Institut Pascal, Clermont-Ferrand, France}

\author{V.~Develay}
\affiliation{Laboratoire Charles Coulomb (L2C), Universit\'e de Montpellier, CNRS, Montpellier, France}

\author{C.~Brimont}
\affiliation{Laboratoire Charles Coulomb (L2C), Universit\'e de Montpellier, CNRS, Montpellier, France}

\author{L.~Doyennette}
\affiliation{Laboratoire Charles Coulomb (L2C), Universit\'e de Montpellier, CNRS, Montpellier, France}

\author{E.~Cambril}
 \affiliation{Centre de Nanosciences et de Nanotechnologies, CNRS, Universit\'e Paris-Saclay, France}
 
 \author{S.~Bouchoule}
 \affiliation{Centre de Nanosciences et de Nanotechnologies, CNRS, Universit\'e Paris-Saclay, France} 

 \author{B.~Alloing}
 \affiliation{UCA, CRHEA-CNRS, Rue Bernard Gregory, 06560 Valbonne, France}
 
 \author{E.~Frayssinet}
 \affiliation{UCA, CRHEA-CNRS, Rue Bernard Gregory, 06560 Valbonne, France}
 
 \author{J.~Z\'u\~niga-P\'erez}
\affiliation{UCA, CRHEA-CNRS, Rue Bernard Gregory, 06560 Valbonne, France}
\affiliation{MajuLab, International Research Laboratory IRL 3654, CNRS, Université Côte d’Azur, Sorbonne Université, National University of Singapore, Nanyang Technological University, Singapore, Singapore}

 \author{T.~Ackemann}
 \affiliation{SUPA and Department of Physics, University of Strathclyde, Glasgow G4 ONG, Scotland, UK}

\author{G.~Malpuech}
 \affiliation{Universit\'e Clermont Auvergne, CNRS, Institut Pascal, Clermont-Ferrand, France}
 
 \author{D.D~Solnyshkov}
\affiliation{Universit\'e Clermont Auvergne, CNRS, Institut Pascal, Clermont-Ferrand, France}
 \affiliation{Institut Universitaire de France (IUF), 75231 Paris, France}
 
 \author{T.~Guillet}
 \email{thierry.guillet@umontpellier.fr}
 \affiliation{Laboratoire Charles Coulomb (L2C), Universit\'e de Montpellier, CNRS, Montpellier, France}

\date{\today}

\begin{abstract}
So far, exciton-polariton (polariton) lasers were mostly single mode lasers based on microcavities. Despite the large repulsive polariton-polariton interaction, pulsed mode-locked polariton laser was never reported.
Here, we use a $60 \ \mu m-$long GaN-based waveguide surrounded by distributed Bragg reflectors forming a multi-mode horizontal cavity. We demonstrate experimentally and theoretically a polariton mode-locked micro-laser operating in the blue-UV, at room temperature,  with a 300$\ GHz$ repetition rate and 100$\ fs$-long pulses. The mode locking is demonstrated by the compensation (linearization) of the mode dispersion by the self-phase modulation induced by the polariton-polariton interaction. It is also supported by the observation in experiment and theory of the typical envelope frequency profile of a bright soliton.
\end{abstract}

%\pacs{}% PACS, the Physics and Astronomy
                             % Classification Scheme.
%\keywords{}%Use showkeys class option if keyword
                              %display desired
\maketitle

% Maximum 3750 words
\section{Introduction} 
Exciton-polaritons (polaritons) are photon modes mixed with excitons. The frequency range around the exciton frequency where this mixing is sizable is a few times the so-called Rabi splitting, which quantifies the strength of the exciton-photon interaction. Their exciton fraction confers polaritons a strongly interacting character. In the language of non-linear optics, this corresponds to a large $\chi^{(3)}$ Kerr-like non-linearity.
This feature was well identified in the 2000's, when parametric amplification~\cite{Saba2001a}, parametric oscillation \cite{savvidis_angle-resonant_2000,baumberg_parametric_2000,stevenson_continuous_2000,Whittaker2001}, and bistability \cite{Baas2004a,Baas2004} have been observed in planar microcavities. The main difference with respect to  $\chi^{(3)}$ processes in other non-linear media is that the resonant processes associated with polaritons are several orders of magnitude more efficient.

These results were obtained in GaAs-based microcavities, where the Rabi splitting is typically $10~meV$ resulting in a relatively narrow bandwidth detrimental to mode-locking, and where excitons are only stable at very low temperature. A first approach to move to room temperature polaritonic devices has been based on the use of large band gap semiconductors, namely GaN and ZnO, for which room temperature polariton lasing has been first proposed \cite{malpuech_room-temperature_2002,zamfirescu_zno_2002} and then observed \cite{christopoulos_room-temperature_2007,baumberg_spontaneous_2008,christmann_room_2008,li_excitonic_2013}. The development of room-temperature polaritonics has been widened thanks to the use of organics~\cite{kena-cohen_room-temperature_2010,plumhof_room-temperature_2013} and, more recently, of perovskites~\cite{Su_Room_2018,Su_Observation_2020,Lu_Engineering_2020, Tao_Halide_2022, Peng_Room_2022}  and transition metal dichalcogenides (TMDs)~\cite{Zhao2021}. However, the smaller exciton Bohr radius and, in most structures, the use of a bulk active medium instead of quantum wells makes the non-linear response in GaN and ZnO weaker than in equivalent GaAs structures. Thus non-linear effects have not been straightforwardly observed in these materials. In parallel to the developments in vertical microcavities, and starting from the middle of the 2010's, the study of polariton waveguides and polariton photonic crystal slabs has gained a lot of interest due to the expected benefits in terms of polariton propagation speeds, polariton lifetimes and ease of fabrication~\cite{Bajoni2009, Suarez-Forero_Enhancement_2021}. Using GaAs polariton waveguides and resonant excitation \cite{walker_exciton_2013}, evidence for bright soliton creation was reported \cite{walker_ultra-low-power_2015}, and {explained by the combination of repulsive polariton-polariton interactions and a positive group velocity dispersion: the associated non-linearity is two orders of magnitude larger than in other standard nonlinear media (for instance, SiN).
In large band gap semiconductor waveguides, polariton lasing was predicted \cite{solnyshkov_optical_2014} and observed recently both in ZnO \cite{jamadi_edge-emitting_2018} and GaN \cite{Souissi_Ridge_2022} horizontal cavities. These in-plane cavities were several tens of micrometers long, opening up the potential for mode-locking via four-wave mixing processes. Indeed, a weak signature of non-linear polariton-polariton interaction under non-resonant optical pumping was reported in a waveguide containing GaN-quantum wells as an active medium \cite{Ciers_Polariton_2020}.

In Fabry-Perot cavities, photonic modes are equally spaced in frequency only if the material is non-dispersive,\textit{ i.e.} if it has a constant group index. However, when a dispersive Kerr-medium fills the cavity, non-linearities renormalize the eigenmode frequencies by an effect called self-phase modulation \cite{Agrawal_Self_1989,Agrawal_Nonlinear_2001} in optics. As a matter of fact, indication for self-phase modulation in a GaN-based planar waveguide has been observed under pulsed resonant excitation \cite{Paola_Ultrafast_2021}. Under these conditions, scattering processes involving two particles at energy $E$ and giving rise to particles at energies $E+\Delta E$ and $E-\Delta E$ become phase-matched and, thus, allowed{, leading for example to the well-studied parametric oscillation in polariton microcavities \cite{baumberg_parametric_2000,stevenson_continuous_2000} and polariton waveguides \cite{Suarez-Forero_Enhancement_2021}}. So-called mode-locking can occur, where instead of a monomode CW laser a bright temporal soliton, with equally spaced frequency components, forms and propagates back and forth in the cavity. This dynamic behaviour results in a pulsed laser output \cite{Hofer_Mode_1991,Vasilev_Fast_2000,Keller_Ultrafast_2010}. 
Mode-locking exploiting the polaritonic nonlinearity has been suggested or explicitely predicted~\cite{jamadi_edge-emitting_2018,Ciers_Polariton_2020,Paola_Ultrafast_2021,egorov_frequency_2018} but it has never been observed so far.

Regarding alternative near-ultraviolet and visible (NUV-VIS) mode-locked lasers, most realizations are presently based on fiber-based systems or frequency conversion. The only demonstrations based on semiconductor laser diodes rely on an external cavity\cite{Weig_Implementation_2014} or a saturable absorber in a multi-section device \cite{Vasilev_Mode_2013} (see Ref.~\cite{Hermans_chip_2022} for a review). More generally, chip-scale advanced NUV-VIS lasers based on the feedback of an absorber or an external resonator mostly rely on the heterogeneous integration of a nitride laser diode and a dielectric photonic resonator \cite{Savchenkov_Self_2019,Wunderer_Single_2023}, presently limiting the advances of nitride-based photonic integrated circuits.

In this work we investigate waveguide polariton lasers based on a GaN ridge waveguide surrounded by in-plane distributed Bragg reflectors, which define horizontal Fabry-Perot resonators. In the 60 $\mu m$-long cavity, we resolve Fabry-Perot modes and determine thereby the corresponding exciton-polariton dispersion. At low temperature (70~$K$) monomode polariton lasing is observed. At higher temperature (from 150~$K$ to room temperature), where polariton relaxation is enhanced, lasing takes place at lower energy and becomes intrinsically multimode. The renormalization of the eigenenergies gives rise to equally spaced mode frequencies, which is a clear evidence of the buildup of resonant parametric processes where non-linear interactions produce self-phase modulation compensating for the polariton group velocity dispersion. We find a polariton mediated Kerr non-linear index $n_2\approx 5\times 10^{-13} \ cm^2 \ W^{-1}$, orders of magnitude larger than non-linearity based on other mechanisms in semiconductors and in full agreement with theoretical expectations based on the magnitude of the polariton-polariton interactions. The envelope of the lasing modes is well fitted by a bright soliton wave function. This picture is confirmed by numerical simulations using the modified Gross-Pitaevskii equation and considering the precise geometry of our cavity. 
 We thus demonstrate the first pulsed polariton laser displaying mode-locking and exploit, unambiguously, polariton non-linearities in large band gap semiconductors. Furthermore, our pulsed polariton laser operates up to room temperature and is based on a compact chip-scale device compatible with electrical injection and photonic integration, promising exploitation in integrated NUV-VIS photonics.

\section{Results}

\subsection{Sample design}
Two similar samples are investigated in this work. The epilayer heterostructure (sample~A) consists of a 3-$\mu m$-thick GaN buffer, a 1.5-$\mu m$-thick Al$_{0.08}$Ga$_{0.92}$N cladding and a 150-$n m$-thick GaN waveguide core grown by metal-organic vapor phase epitaxy (MOVPE) on c-plane sapphire; the sample~B differs by an additional Al$_{0.08}$Ga$_{0.92}$N cap layer of 20~$n m$ intended to improve the carrier confinement at room temperature (see Appendix~\ref{app:Multimode lasing at room temperature}, Fig.~\ref{fig:samples}). The planar slab waveguide is then patterned by electron beam lithography and dry deep etching, as described in Ref.~\cite{Souissi_Ridge_2022} and shown in Figs.~\ref{fig:sample}(a) and ~\ref{fig:sample}(b). The laser devices consist of 1~$\mu m$-wide ridge waveguides ended by 4~pairs in-plane distributed Bragg reflectors (GaN-air DBRs). This work focuses on a device with a cavity length $L_\text{cav}=60~\mu m$.

\begin{figure}[b]%[tbhp]
\centering{\includegraphics[width=8cm]{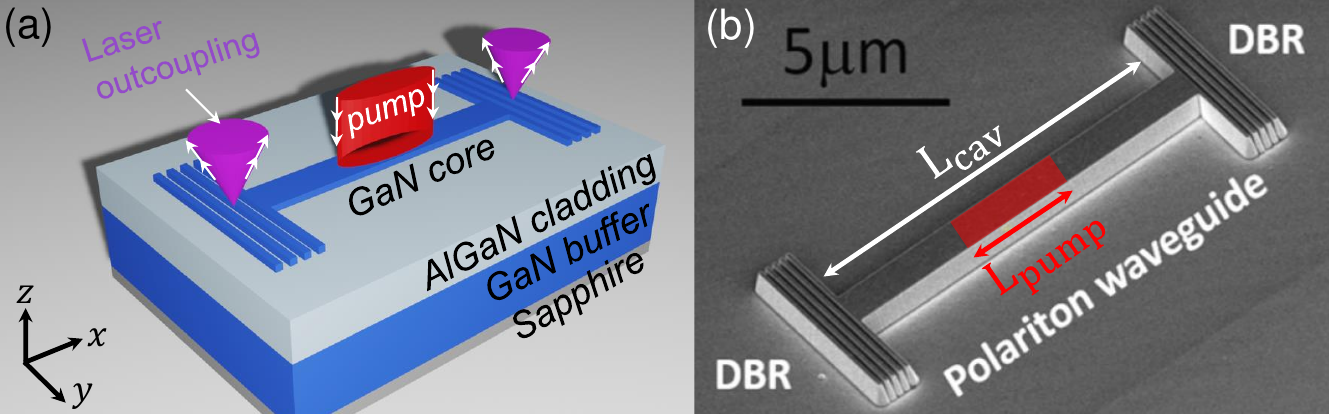}} 
  \caption{Sample description. (a)~Sample structure: ridge cavity after lithography and deep etching. The guided mode propagates along the $x$-axis in the GaN core. (b)~Scanning electron micrograph (SEM) image of a 10~$\mu m$-long polariton ridge waveguide in sample~A. A  10~$\mu m$-long device is shown instead of the experimentally used 60~$\mu m$-long device to have enough resolution to resolve the Bragg mirror structure. }
  \label{fig:sample}
\end{figure}

\subsection{Linear regime: polariton dispersion below threshold}
The in-plane cavities are optically pumped with a ns-pulsed laser $(355\ nm,$ $4\ ns$ pulses, see Appendix~\ref{app:Cavity imaging} for additional details) that excites resonantly the exciton reservoir, as sketched in Fig.~\ref{fig:dispersion}(a). The line-shaped pump profile (Fig.~\ref{fig:sample}) is adjusted to excite partially, or entirely, the cavity length (pump length $L_\text{pump})$, controlling thereby the size of the exciton reservoir. These excitons in the reservoir relax and feed the lower polariton branch (LPB). The polariton dispersion evidences the strong exciton-photon coupling regime through the anti-crossing of the LPB and the exciton resonance (Fig.~\ref{fig:dispersion}(a)), as measured and modelled in similar planar waveguides in Ref.~\cite{Brimont_Strong_2020}. Within the cavity, the LPB is quantized into Fabry-Perot modes with discrete wavevectors, whose signal is partially redirected to the vertical direction by the DBRs, which act not only as in-plane mirrors but also as outcouplers. This signal can be collected by the microscope objective of a micro-photoluminescence set-up. Figure~\ref{fig:dispersion}(c) illustrates the emission spectrum of a $60 \mu m$-long cavity in linear scale. The measurement was performed at a temperature of 150~$K$ with a power level significantly below threshold. The spectrum displays Fabry-Perot modes from $3.380$ to $3.435\ eV$. 

The strong coupling regime between photons and excitons is assessed through the modeling of the measured cavity free spectral range (FSR), which is proportional to the first derivative of the polariton energy dispersion ($E_\text{LPB}$), {\it i.e.}, to the group velocity $v_\text{g}$ (see Appendix~\ref{app:Strong coupling regime and group velocity dispersion}, eq.~\eqref{FSR}). Figure~\ref{fig:dispersion}(d) shows a good quantitative agreement when considering the strong-coupling between the photonic TE0 mode and the GaN excitons, following the procedure detailed in Ref.~\cite{jamadi_edge-emitting_2018,Souissi_Ridge_2022}.
The decrease of the FSR from 2.0 to $0.9 \ meV$ (55 $\%$ change) is reproduced by the polariton dispersion (solid red line), exhibiting a strong group velocity dispersion (GVD, see Appendix~\ref{app:Strong coupling regime and group velocity dispersion}). On the other hand, the FSR calculated without exciton resonances in the dielectric function (``TE0 bare photon'', dashed black lines) is almost constant around $2.2 \ meV$ varying by only 12 $\%$ in the same energy range. In the model the GaN exciton energy $E_\text{XA} = 3.475 \ eV$ is determined from the reflectivity spectra of the same active layer taken at 150~$K$, whereas the core and cladding thicknesses are measured from scanning electron micrographs. At zero exciton-photon detuning $(\delta = 0)$, this model provides a precise estimate of the Rabi splitting, $\Omega_\text{Rabi} = 64 \pm 10 \ meV$, characterizing the exciton-photon coupling strength, as shown in Fig.~\ref{fig:dispersion}(a). 

\begin{figure}[h]%[tbhp]
\centering{\includegraphics[width=8cm]{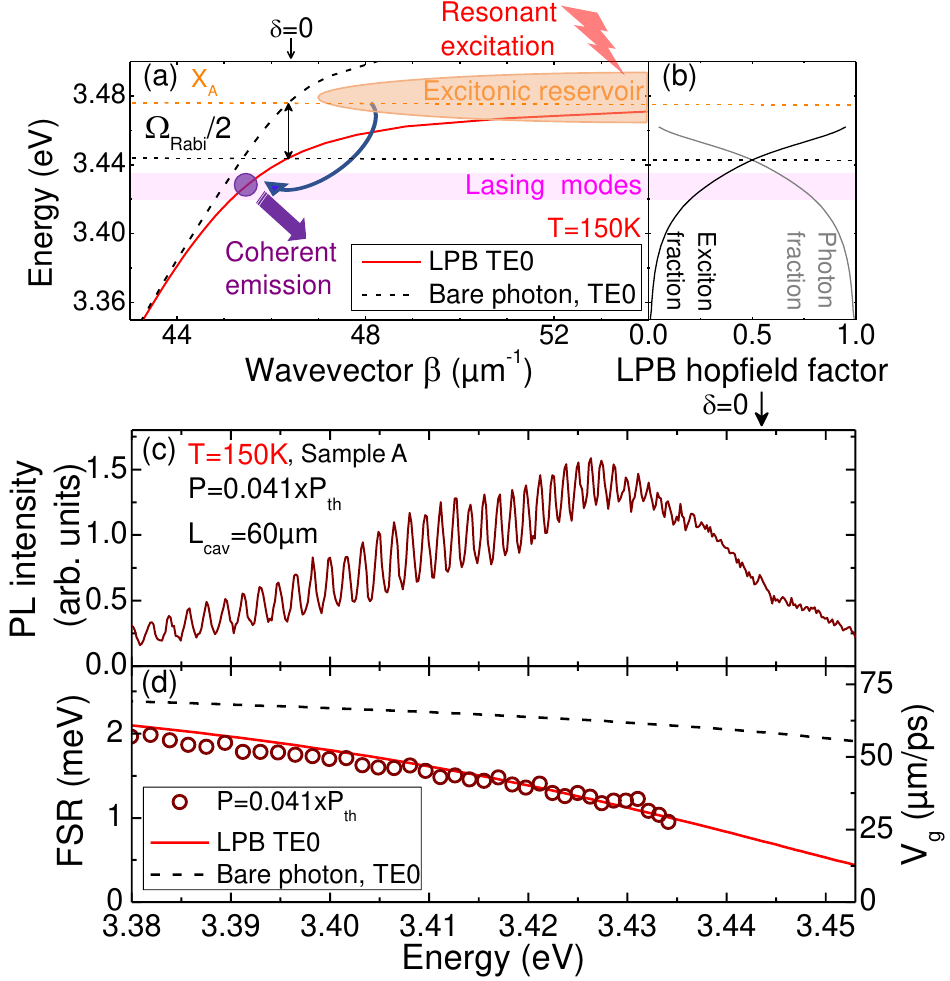}} 
  \caption{(a)~Waveguide polariton dispersion and scheme of the mechanisms involved in the polaritonic lasing; (b)~Excitonic and photonic Hopfield coefficients of the LPB; (c)~Emission spectrum of a 60~$\mu$m-long cavity collected at power $P=0.041\times P_\text{th}$ taken at 150$\ K$ in sample~A; (d)~Experimental FSR (dots) and calculated FSR from LPB dispersion (solid red line) and from the dispersion of the bare photon without excitons (dashed black line){, for the TE0 mode of a planar waveguide}.}
  \label{fig:dispersion}
\end{figure}

\subsection{Nonlinear regime at 70~K: single-mode polariton laser operation}
\label{par:Nonlinear regime at 70 K: single-mode polariton laser operation}

In a waveguide polariton laser, lasing occurs as soon as the population of one of the Fabry-Perot cavity modes overcomes unity so that bosonic stimulation enhances the scattering processes towards the lasing mode. Polariton laser operation in a continuous wave (CW) regime has been demonstrated for the investigated devices in Ref.~\cite{Souissi_Ridge_2022}, where they were modelled in detail.  It was notably shown that polariton lasing could be achieved by pumping the entire laser cavity, as in a standard edge-emitting laser, but also when pumping only a fraction of the total length, which can be much smaller than $50\%$ of the cavity length. This establishes a clear distinction between waveguide polariton lasers and edge-emitting lasers based on an electron-hole plasma. Furthermore, it was shown that when the pump length $L_\text{pump}$ amounts to only $15\%$ of the cavity length, the polariton laser threshold shows just a three-fold increase compared to pumping the whole cavity.

Figure~\ref{fig:Single_Multimode}(a) presents the CW laser operation measured at $\text{T}=70 \ K$, which is very similar to the one reported in Ref.~\cite{Souissi_Ridge_2022}: the spectrally-resolved emission below threshold shows a set of Fabry-Perot modes covering a wide spectral range. Upon increasing the pumping intensity, some modes around 3.458~$eV$ (close to zero exciton-photon detuning $(\delta \approx 0)$) undergo a strong nonlinear intensity increase and a line narrowing, indicating {quasi-single-mode} CW lasing that rapidly extends to a few modes \cite{Yang_Topological_2022}. Single-mode CW lasing for a shorter pump length ($\ L_\text{pump}= 6.5 \ \mu m$) is presented in Appendix~\ref {app:CW lasing at T=70K} (Fig.~\ref{fig:Monomode_small_spot}) for exactly the same cavity. 

\subsection{Nonlinear regime from 150~K to room-temperature: multimode polariton laser operation with mode synchronization}

\begin{figure*}
\centering{\includegraphics[width=10cm]{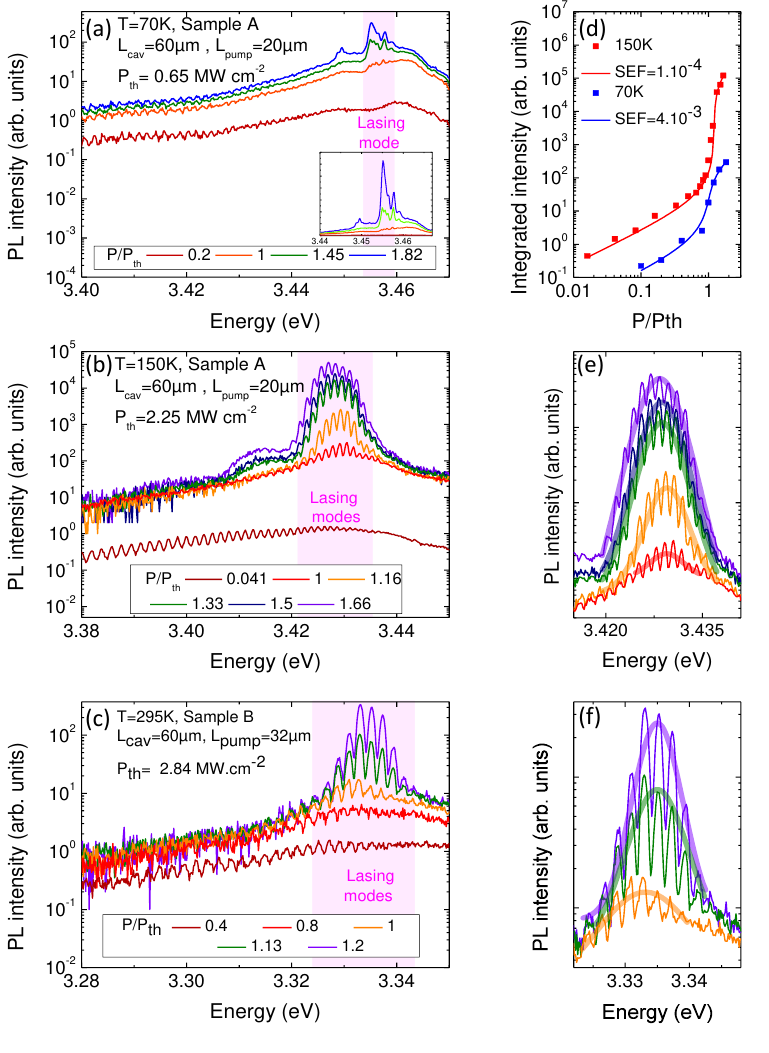}} 
  \caption{
  Polariton laser operation of a 60~$\mu m$-long cavity. Power-dependent emission spectra across the lasing threshold, at $\text{T}=70 \ K$ (a)~(inset shows the intensity blow-up across the lasing threshold in linear scale), $\text{T}=150 \ K$ (b) and $\text{T}=295 \ K$ (c); (d)~Corresponding peak integrated intensity (square dots) and modelling of the spontaneous emission factor $SEF$ (solid line) at  $\text{T}=70 \ K$ (in blue) and $\text{T}=150 \ K$ (in red); (e,f)~Fit by a squared hyperbolic secant lineshape (thick light-colored plain lines, corresponding to Eq.~\eqref{eq:sech}) of the envelope of the emission spectrum above threshold (thin plain lines).}
  \label{fig:Single_Multimode}
\end{figure*}

As the sample temperature is increased to $\text{T}=150 \ K$ or above, in particular at room temperature, (Fig.~\ref{fig:Single_Multimode}(b,c)), the spectrum is almost unchanged below threshold. However, for the same pump length as in section~\ref{par:Nonlinear regime at 70 K: single-mode polariton laser operation}, the threshold value is 3.5 times and 4.5 times larger than at $70 \ K$, respectively. The behaviour at threshold is radically different since the emission is inherently multimode, with a well-defined envelope nicely reproduced by a secant hyperbolic function (Fig.~\ref{fig:Single_Multimode}(e,f)), see Appendix~\ref{app:Secant hyperbolic pulse propagation in a dispersive nonlinear material}), and about ten modes involved in the non-linear emission. The Hopfield coefficients of the lasing modes are about $25\%$ excitonic and $75\%$ photonic (Fig.~\ref{fig:dispersion}(b)) at $\text{T}=150 \ K$ and become more photonic at $\text{T}=295 \ K$ ($10\%$ excitonic, $90\%$ photonic). The input-output characteristics (Fig.~\ref{fig:Single_Multimode}(d)) show that the non-linear increase of the emitted intensity is much steeper at $\text{T}=150 \ K$ (multimode case) than at $\text{T}=70 \ K$ (monomode case), with a spontaneous emission factor SEF decreasing by a factor 40, from $4\times 10^{-3}$ to $1\times 10^{-4}$. 

The change of the waveguide polariton dispersion at threshold provides a detailed insight into the phase synchronization of the laser modes. We therefore analyze the energy shift of each cavity mode across threshold  (Fig.~\ref{fig:Weakening_fosc}(b)) and the corresponding cavity FSR (Fig.~\ref{fig:FSR}(a)). The spectra at $\text{T}=150 \ K$ are chosen for this detailed analysis since they present the largest mode contrast and the broadest series of modes for a largest pump power range. The same analysis at $\text{T}=295 \ K$ is provided in the Appendix~\ref{app:Multimode lasing at room temperature} (Fig.~\ref{fig:FSR_295K}).

Beyond threshold, it is noteworthy that the FSR of the lasing modes around $3.43~eV$ becomes constant (Fig.~\ref{fig:FSR}(a)), indicating that these modes now propagate at a common group velocity (right axis). The number of modes with a constant FSR increases from 6~modes at lasing threshold  $P_\text{th}$, to 10~modes at $P=1.66\times P_\text{th}$ (Fig.~\ref{fig:FSR}(c)). A more detailed sweep in terms of pumping intensities can be found in Fig.~\ref{fig:FSR_waterfall}. The corresponding linearization of the polariton dispersion, deduced from the measured energy shift of each lasing mode, is shown in Fig.~\ref{fig:FSR}(b) (bare data in  Figs.~\ref{fig:Weakening_fosc}(a) and ~\ref{fig:Weakening_fosc}(b)). The data are directly deduced from Fig.~\ref{fig:Weakening_fosc}(b) by attributing the value $\beta_0 = 45.3 \ \mu m^{-1}$ to the central lasing mode at $3.423~eV$, in accordance with the theoretical dispersion far below threshold, and then adding multiples of $\pm \ \pi / L_\text{cav}$ to $\beta_0,$ as imposed by the wavevector quantization in the cavity. 

\begin{figure} [b] %[tbhp]
\centering{\includegraphics[width=7cm]{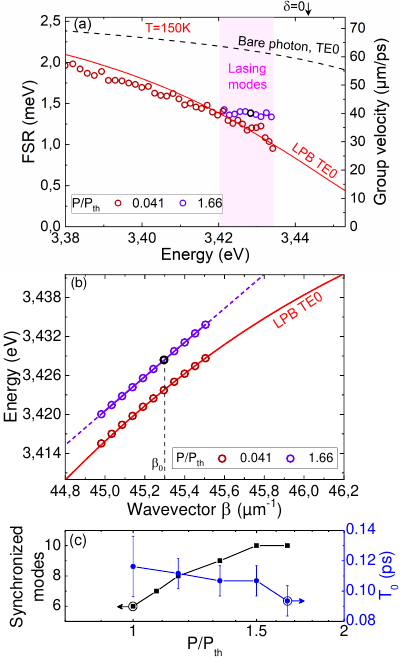}} 
  \caption{Free Spectral Range (FSR) and group velocity $v_\text{g}$ of the modes across the lasing threshold {for the sample A}, at $\text{T}=150 \ K$ (a) and the corresponding evolution of the waveguide polariton dispersion {calculated for the TE0 mode of the planar waveguide} (b). Polariton dispersions below and above threshold (respectively red and violet). {The calculated dispersion for the TE0 mode is shown as lines; the circles correspond to the longitudinal modes quantized in the $60 \ \mu m$-long Fabry-Perot cavity. The black open circle indicates the central lasing mode at wavevector $\beta_0$}. The dispersion beyond threshold (violet) is linear, and extended with a dotted line as a guide for the eye. (c)~Number of synchronized modes and temporal pulse width $T_0$ deduced from the secant hyperbolic fit of the spectra envelope.  
  }
  \label{fig:FSR}
\end{figure}

In semiconductor edge-emitting lasers operating in the weak coupling regime, the linearization is usually interpreted in terms of mode-locking, which occurs when the group velocity dispersion parameter $\beta_2$ and the non-linear refractive index $n_2$ (defined in Appendix~\ref{app:Strong coupling regime and group velocity dispersion} and \ref{app:Nonlinear refractive index and polariton-polariton interaction constant}) have opposite signs and the self-focusing condition is fulfilled \cite{Agrawal_Nonlinear_2001}. In exciton-polariton systems, the non-linear refractive index is linked with a well-defined microscopic mechanism: polariton-polariton interactions. Meanwhile, the modification of the dispersion is interpreted in terms of the elementary excitations of the polariton fluid, at the heart of parametric processes \cite{savvidis_angle-resonant_2000,savvidis_asymmetric_2000} and superfluidity \cite{Utsunomiya_Observation_2008, amo_polariton_2011}.

The spectral linewidth of the pulses (Fig.~\ref{fig:Single_Multimode}(e)) increases with the pumping power beyond the laser threshold, and the corresponding duration of the pulses $T_0$ decreases from $110 \ fs$ to $90 \ fs$ (Fig.~\ref{fig:FSR}(c)). The pulse width at half maximum $T_\text{FWHM}$ (see Appendix~\ref{app:Secant hyperbolic pulse propagation in a dispersive nonlinear material}, Eq.~\eqref{FWHM}) is about 17~times smaller than the period between pulses $(3 \ ps$ for $L_\text{cav}=60 \ \mu m$ and $v_g \approx 40 \ \mu m.\ ps^{-1}).$ 

Note that the waveguide polariton lasers discussed so far in this work were achieved by pumping just a fraction of the cavity length, similar to our previous CW operation demonstration \cite{Souissi_Ridge_2022}. One can wonder what role is actually played by the unpumped section of the cavity. Figure~\ref{fig:dependence on the pump length} presents the spectra of the same device discussed above but pumped along almost the full cavity length $(L_\text{pump}/L_\text{cav} = 0.83$ instead of $0.33$). The effect on the laser threshold is very similar to the one reported under CW operation conditions\cite{Souissi_Ridge_2022}: when increasing the length of the pumping spot by a factor $2.5$, the threshold pump power density decreases by a factor 3. The threshold pump energy per pulse, integrated over the pumped section, increases only by $20\%,$  from $1.5~nJ/pulse$ to $1.8~nJ/pulse$. This attests to the small impact of the potential additional losses associated with a large unpumped section in the cavity, which would have been totally detrimental in an edge-emitting laser operating in the weak coupling regime. Interestingly, the mode-locking dynamics is independent of the pump length: it is characterized by the same lasing energy, the same multimode emission with a hyperbolic secant spectral lineshape, and thus the same deduced pulse widths $T_0$. This demonstrates that the unpumped section of the cavity does not act as a saturable absorber.

\subsection{Simulation of polariton mode-locking dynamics}

In order to understand the lasing regimes observed in the experimental results, we simulate the laser dynamics inside the waveguide cavities thanks to the temporal resolution of the 1D Gross-Pitaevskii equation (GPE) for coupled excitons and photons, which is closely related to the non-linear Schrödinger equation describing mode locking in lasers. In general, theoretical modelling of the coupled condensate-reservoir system always needs to focus on the detailed description of one of the subsystems: the condensate or the reservoir. This is due to the opposite nature of the two: the condensate is fully coherent (well-described with the Gross-Pitaevskii equation), and the reservoir is fully incoherent (well-described with the Boltzmann equations). The choice depends on what is more important in the particular problem: the dynamics of the condensate \cite{wertz_propagation_2012,Solnyshkov2014pra} (in which case the reservoir needs to be simplified) or the dynamics of the reservoir\cite{solnyshkov_optical_2014,Ciers_Polariton_2020,Souissi_Ridge_2022} (in which case the condensate coherent dynamics is neglected). To describe the mode locking due to the interactions in the condensate, we obviously need to focus on the dynamics of the condensate, therefore neglecting the reservoir. The equations, qualitatively similar to \cite{wertz_propagation_2012}, read:
\begin{equation}
i\hbar\frac{\partial \psi}{\partial t} = \hat{T}\psi+U_1\psi +\Omega_R\chi/2
\label{GPE1}
\end{equation}

\begin{multline}
 i\hbar\frac{\partial \chi}{\partial t}=i\gamma e^{-n_x/n_{max}}R(x)\hat{\Gamma}\chi +\xi(x,t)+\alpha_{x,1D}|\chi|^2\chi \\ +U_2\chi+E_{XA}\chi+\Omega_R\psi/2
 \label{GPE2}
\end{multline}
where $\psi$ is the photon wavefunction, $\chi$ the exciton wave function, $\Omega_\text{R}$ the Rabi splitting value. The kinetic energy operator $\hat{T}$ encodes the 1D-dispersion of the photonic guided modes and is defined as $\hat{T}\psi=F^{-1}\left(E(k)F(\psi)\right)$,
with $F$ being the 1D Fourier transform, $F^{-1}$ its inverse, and $E(k)$ the dispersion of the guided photonic mode. Similarly, the gain operator $\hat{\Gamma}$ is defined by $\hat{\Gamma}\chi=F^{-1}\left(\Gamma(k)F(\chi)\right)$, where 
$\Gamma(k)$ is the gain profile, which we use as an input of the model respecting features obtained in previous works \cite{solnyshkov_optical_2014,Ciers_Polariton_2020,Souissi_Ridge_2022}. Qualitatively the gain profile shifts toward lower energy and becomes wider when increasing either the temperature or the pumping power.  The coefficient $\gamma$ is a scattering rate, $n_\text{x}=L_\text{cav}^{-1}\int|\chi|^2\,dx$ is the mean  exciton density in the condensate and $n_\text{max}$ a saturation density for the mean value of the 1D density (the density can locally be larger). 
$\xi(x,t)$ is a weak white noise term aiming to describe spontaneous scattering processes, which are absent from the Gross-Pitaevskii equation. $U_1$ and $U_2$ are the photonic and excitonic potentials, respectively, both containing high barriers at the end of the cavities with an imaginary part representing the losses at the DBRs. 
The excitonic potential $U_2$ also contains a contribution of the exciton reservoir $U_\text{R} R(x)$, with a variable amplitude $U_\text{R}$, whose profile is defined by the pumping spatial distribution $R(x)$. This reservoir contribution embeds both the exciton-exciton interaction and the screening of the exciton oscillator strength. $\alpha_{X,1D}$ is the 1D exciton-exciton interaction constant (see Appendix~\ref{app:Nonlinear refractive index and polariton-polariton interaction constant}) at the heart of the $\chi^{(3)}$ polariton-polariton non-linearity.
\begin{figure*}%[tbhp]
\centering{\includegraphics[width=12cm]{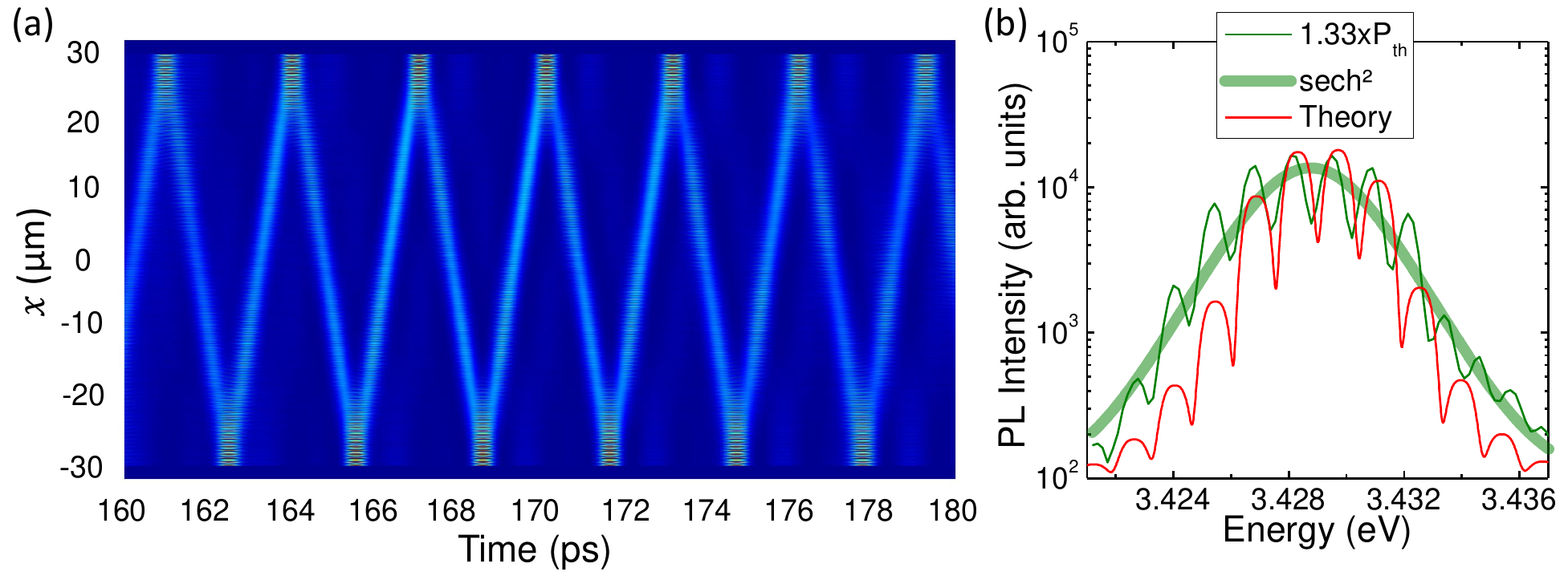}}
  \caption{(a)~Simulation of the time- and spatially-resolved polariton population after the start  of a continuous pumping of the exciton reservoir at $t=0$ (for $\alpha n_\text{max}=0.02 \ meV$); $L_\text{cav} = 60 \ \mu m$ , $L_\text{pump} = 20 \ \mu m,$ $ \text{T}=150 \ K$. (b)~Comparison between the output spectra measured at $P=1.33\times P_\text{th}$  (thin green line { (Sample A)}) and calculated (thin red line); Hyperbolic secant squared fit (thick green line).}
  \label{fig:bright_soliton}
\end{figure*}

To summarize, the adjustable parameters of the model are the gain profile $\Gamma(k)$ (in particular, its width) and the saturation density $n_\text{max}$. All other quantities (including the center of the gain profile) are extracted from the experiment or calculated from first principles. The simulation is performed by solving the equations~\eqref{GPE1} and \eqref{GPE2} over time with zero initial condition, until the system settles in a quasistationary regime (see Appendix~\ref{app:Numerical simulations} for additional details).

For a narrow gain profile ($2~meV$ wide) centered at polariton states corresponding to a detuning $\delta=-5~meV$ and weak interactions $\alpha_{x,1D} n_\text{max}\approx 0.003~meV$, the model predicts a stationary monomode lasing regime in agreement with experimental observations made at $70 \ K$ and in \cite{Souissi_Ridge_2022}.
To describe data taken at $150 \ K$, we use a gain centered at detuning $\delta=-20~meV$, $\alpha_{x,1D} n_\text{max}\approx 0.02~meV$ and we adjust the gain width to reproduce the experiment. Figure~\ref{fig:bright_soliton}(a) shows the results of this simulation for a gain width of $6~meV$, for which a single bright temporal soliton forms in the cavity, with a roundtrip period of $3.05~ps$. 

Figure~\ref{fig:bright_soliton}(b) shows the Fourier transform of the temporal evolution of the soliton over a time window of $3~ns$ taken at one edge of the waveguide. It is compared with the experimentally measured output spectrum at $P=1.33\times P_\text{th}$. The spacing between peaks in theory is almost constant and equal to $1.35~meV$ for all the ten modes shown in the figure, in good agreement with the experiment. On the other hand, the frequency width of the envelope is smaller by about $20\%$ in theory. The theoretical spectrum can be made broader using a wider gain. However, this leads to the development of several temporal solitons competing within the cavity and the spectrum becomes less regular. We cannot conclude currently if this is a feature of the simplified model we use (we neglect the decoherence and energy relaxation within the polariton modes, for example), or if this multi-soliton regime is also present in the experiment. The experimental spectrum seems very regular and is well fitted by a $\sech^2$ function, which tends to suggest a single soliton regime. The simulation on Fig.~\ref{fig:bright_soliton} can be compared to the $\sech^2$ profile both in time and in frequency, leading to a time-bandwidth product equal to $0.43$; this value slightly exceeds the $0.315$ lower limit expected for Fourier-transform $\sech^2$ pulses, indicating that a small contribution of chirping exists in the simulated pulses.

\begin{figure}[tbhp]%[!htbp]  
\centering{\includegraphics[width=8cm]{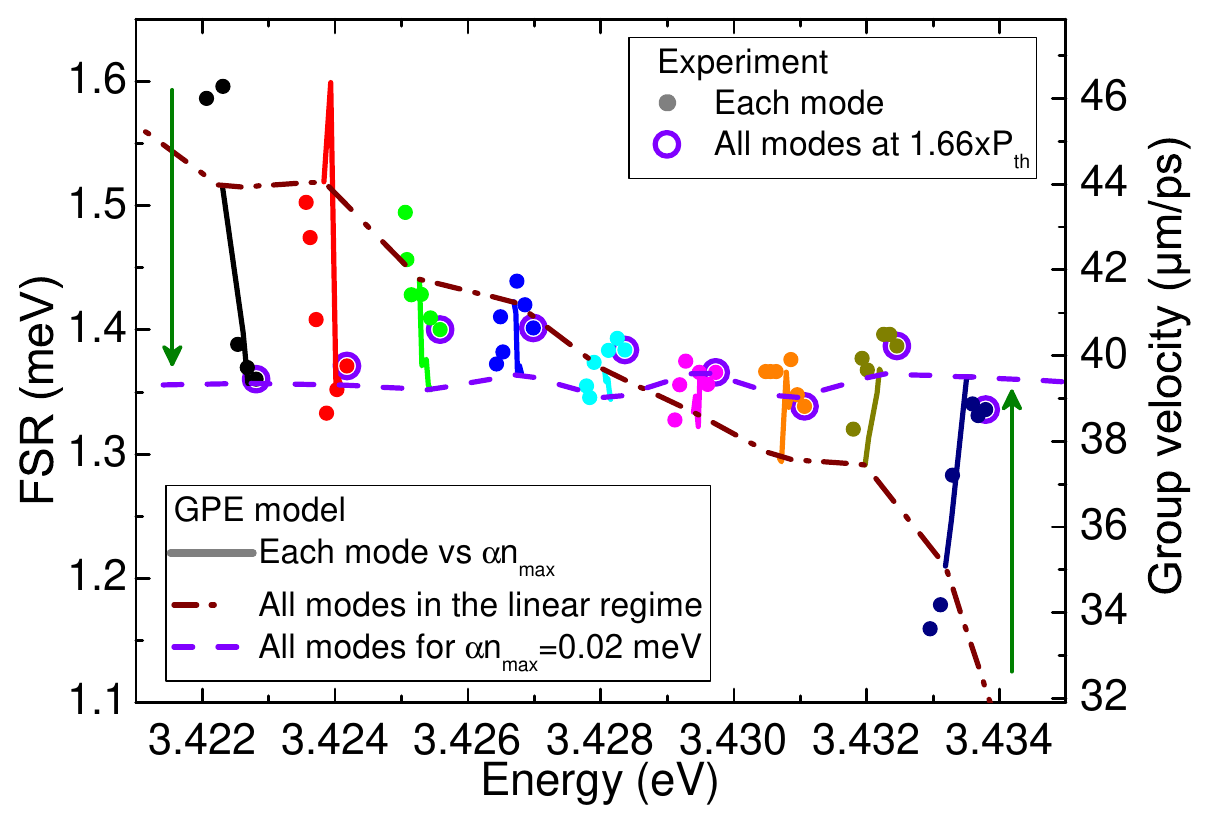}}  

  \caption{
  Trajectory of each lasing mode (dots: one color per mode) in FSR-Energy coordinates for different pumping powers. Data points indicate experimental results {(Sample A)}; the plain lines represent the results of the GPE model for polariton mode locking. Theoretical trajectories of modes in linear regime and for $\alpha n_\text{max}=0.02 \ meV$ are indicated as dashed dot line and as dashed line, respectively.}
  \label{fig:Exp_nonlinearity}
\end{figure}

Figure~\ref{fig:Exp_nonlinearity} shows in solid line, for the same gain width used to simulate Fig.~\ref{fig:bright_soliton}, the theoretically computed FSR for increasing values of pumping (represented by $n_\text{max}$ in the model). It is compared to the experimentally extracted FSR for pumping values ranging between $0.83\times P_\text{th}$ and $1.66\times P_\text{th}$. Below threshold, the experimental FSR (brown dotted-dashed line) varies from $1.55~meV$ to $1.15~meV$ in this frequency range ($20\% $ change), which is also the case in theory, thus confirming that the experimental and theoretical energy dispersion relations are in good agreement. Increasing pumping, the FSR of modes at lower energy decreases from $1.55$ to $1.35~meV$, whereas the FSR of modes at higher energy increases from $1.15$ to $1.35~meV$. The FSR of central modes stays constant. At the highest pumping value the FSR of the 10 modes involved in the soliton becomes constant in energy at $1.35$-$1.36~meV$, within an accuracy of $2\%$, which is a very clear signature of a mode-locking effect. The theoretical calculation reproduces this effect completely. The trajectory of individual FSR is well described, including the fact that FSR does not evolve monotonically versus pumping: FSR can slightly increase first and then decrease towards $1.35~meV$, or viceversa.

These observations clearly confirm the possibility of mode-locking and self-pulsing via polaritonic non-linearities as suggested in earlier works on polariton amplification in ZnO and GaN polariton waveguides \cite{jamadi_edge-emitting_2018,Ciers_Polariton_2020,Paola_Ultrafast_2021,Delphan_Polariton_2023}, and in theoretical proposals on frequency comb generation in polaritonic microring resonators \cite{egorov_frequency_2018}, but had never been achieved.

\section{Quantification of polaritons non-linearities }

The observed mode-locking of a polariton laser exhibits distinct characteristics compared to the standard figures of merit in other non-linear optical systems (as defined in \cite{Agrawal_Nonlinear_2001} and detailed in the Appendix) : 

(i) Due to the short cavity length, the roundtrip time of the pulse train ($3~ps$) corresponds to a repetition frequency larger than $300 \ GHz$. A direct time-resolved observation of the pulsed regime is therefore very difficult to provide experimentally.
 
(ii) As shown in Fig.~\ref{fig:FSR}(a), the group velocity varies by more than $20\%$ (from $40$ to $26 \ \mu m . ps^{-1}$) within the spectral range involved in mode-locked lasing; this relates to a group velocity dispersion parameter $\beta_2 = 4\times 10^{-22} \ s^2 . m^{-1},$ which is 4~orders of magnitude larger than in non-linear optical fibers \cite{Dudley_Supercontinuum_2006} and 2~orders of magnitude larger than in AlGaAs waveguides investigated for soliton formation \cite{belanger_solitonlike_1997}. The corresponding ``dispersion length'' is $L_\text{D} = T_0^2/| \beta_2 | \approx 25 \ \mu m$ \cite{Agrawal_Nonlinear_2001} (see Appendix~\ref{app:Secant hyperbolic pulse propagation in a dispersive nonlinear material}), one order of magnitude smaller than the typical mm-scaled values in non-linear waveguides \cite{belanger_solitonlike_1997} and comparable to the cavity length in the presently investigated cavities.

(iii) The energy per pulse of the current mode-locked laser is difficult to estimate since the in-plane output of the laser devices cannot be directly accessed: indeed, only a small and unknown fraction of the laser emission is collected thanks to the scattering by the DBRs. To obtain a quantitative estimation, we here consider that the lasing modes occupancy at threshold is close to one. The integrated intensity is rising at threshold by a factor $10^4$  at $\text{T}=150 \ K$ (Fig.~\ref{fig:Single_Multimode}(d)). We therefore consider that the soliton contains $N\approx \ 10^4$ polaritons. This corresponds to a soliton  energy of $5.5 \ fJ$, and a soliton peak power $P_0 \approx 55\ mW.$ Such pulse energies are 2 to 3 orders of magnitude smaller than the threshold for phase modulation reported in similar waveguides under resonant excitation $(50pJ$ per $100 \ fs-$long pulse, with a 5 times wider transverse mode profile)\cite{Paola_Ultrafast_2021}. This energy analysis allows estimating which fraction of the pumping laser is transferred to the soliton, which turns out to be of the order of $0.3\%$: indeed, each $4~ns$-long pump pulse has an energy of $1.8~nJ$ at threshold, and generates a train of about 1000 mode-locked pulses, based on the round-trip time. This $0.3\%$ conversion from the optical pump to the polariton laser emission accounts for the geometrical matching between the pump spot and the $1~\mu$m-wide ridge, for the absorption by the active layer, and for the competition between exciton spontaneous emission in the reservoir and exciton stimulated relaxation towards the polariton lasing modes.
 
(iv) The Kerr-type non-linearities involve the polariton-polariton interaction constant $\alpha$ (see Appendix~\ref{app:Nonlinear refractive index and polariton-polariton interaction constant}). The value deduced from the measured FSR shift $(0.2 \ meV)$ through eq.~\eqref{FSR-N} is $\alpha \approx 6.3 \ 10^{-6} meV$, quite close (just a factor 2) to the theoretical value $\alpha \approx 2.5 \ 10^{-6} meV$ when considering exciton-exciton interactions in a waveguide (eq.~\eqref{alpha}). One should notice that in theory the interaction energy in the soliton is 15 times larger than $\alpha_\text{X,1D}n_\text{max}$ because the $N$ polaritons present in the system are not distributed along the 60 $\mu m$ of the sample, but concentrated in the $4 \ \mu m$ healing length of the soliton. We can thereby deduce the non-linear refractive index $n_2$, with an experimental value of $n_2 \approx 4.7 \times 10^{-13} \ cm^2.W^{-1}$ and a  theoretical value just a factor two larger $n_2 \approx  10^{-12} \ cm^2.W^{-1}$, for a polariton-polariton $\chi^{(3)}$ non-linearity. It is three orders of magnitude larger than in bulk non-linear materials, and three times larger than in AlGaAs waveguides \cite{belanger_solitonlike_1997}. Accounting for the transverse effective area of the the guided modes $A_\text{eff}$, we can estimate the non-linear length $L_\text{NL} = n_0 A_\text{eff} / (\beta_0 \ | n_2 | \ P_0) \approx 25-250~\mu m$.

The estimated value of $n_2$ is close to the estimates of ref.~\cite{Paola_Ultrafast_2021} for QW based nitride polariton waveguides, which are similar to ours but without Fabry-Perot cavities. Indeed, in ref.~\cite{Paola_Ultrafast_2021} the deposited gratings are only used for the coupling of the resonant excitation laser to the waveguide polariton modes and their collection. Instead, in the present work the  DBRs define an in-plane cavity providing the necessary laser feedback. Besides, they also outcouple a fraction of the laser emission to the vertical direction, enabling us to collect it. In reference~\cite{Paola_Ultrafast_2021} a spectral broadening of the injected $100~fs$~-~long light pulses was observed and interpreted in terms of polariton non-linearities involving the exciton-exciton repulsion and the reduction of the exciton oscillator strength, both contributing to a density-dependent blueshift of the polariton frequency. The non-linearity was measured for free propagating polaritons without feedback, down to propagation lengths of $100$~$\mu m$ and resonant pump energies of $50~pJ/pulse$. Under non-resonant excitation, similar waveguides without cavity feedback also evidenced the onset of modulation instability of the polariton relaxation~\cite{Ciers_Polariton_2020}, which is an additional evidence of Kerr-like parametric processes.

The present results, with both a flattening of the FSR and a hyperbolic secant spectral lineshape, provide a clear demonstration that the parametric processes foster mode synchronization among the lasing modes. We emphasize that the orders of magnitude of the non-linear length $L_\text{NL}$, of the dispersion length $L_D$ and of the roundtrip length $2 L_\text{cav}$ in our cavities are comparable, which is consistent with the onset of mode-locking in our devices. This exemplifies how polariton waveguide devices are prone to reduce both the length scale and the energy scale of non-linear photonic devices, and increase the operation frequency.

\section{Connection with conventional mode-locked semi-conductors and solid-state lasers}

Passive mode-locking,\textit{ i.e.}\ mode-locking which does not demand an external modulation, is usually achieved via a saturable absorber, most often on semiconductor basis \cite{Keller_Ultrafast_2010}, although the resulting pulse might be close to a Kerr-soliton \cite{haus75a,jung95,paschotta02}. Even in the case of the so-called Kerr-lens mode-locking in Ti:Saphire lasers \cite{spence91} (and some vertical-cavity semiconductor lasers \cite{kornaszewski12,albrecht13}), the mode-locking is primarily not a dispersive effect but achieved via saturable losses,\textit{ i.e.}~a generalized form of saturated absorption, as the change of modal size due to non-linear lensing changes the losses at intra-cavity apertures \cite{Keller_Ultrafast_2010}. Similarly, non-linear polarization rotation in fibre lasers controls losses in the polarization sensitive elements \cite{matsas92}. In edge-emitting semiconductor lasers, saturable absorption is usually provided by an unpumped absorber section and a pumped gain section in so-called two-section devices \cite{Vasilev_Fast_2000,marsh17,wei22,yadav23}, which were implemented in nitride QW-based laser diodes operating in a Q-switch regime \cite{Miyajima_Picosecond_2009,Scheibenzuber_Bias_2010} and a mode-locked regime \cite{Vasilev_Mode_2013}. On the contrary, in our results the independence of the mode-locking on the fraction of pumped cavity length  indicates that the polariton system does not operate as a two-sections device, in the sense described before.

There have been some observations of passive mode-locking in single-section edge-emitting lasers, in particular for quantum dash and quantum dot gain materials \cite{renaudier05,gosset06,rosales12}. This has been sometimes referred to as `magic' mode-locking \cite{liu18}, in analogy to the `magic' mode-locking in Ti:Saphire lasers before the role of saturable losses at apertures in Kerr-lens mode-locking was understood \cite{Keller_Ultrafast_2010}. There has been some debate whether mode-locking in single-section devices is due to intrinsic non-linearities as the Kerr-non-linearity from four-wave-mixing only or whether unintentional saturable absorption in not sufficiently pumped perimeter areas plays a role \cite{chow20}. Recent work indicates that  four-wave mixing in combination with other non-linearities as gain saturation and spatial hole burning can be indeed enough to explain sustained mode-locking \cite{chow20,grillot22,yadav23} but its quality and stability depends strongly on injection levels \cite{chow20} and/or extra-cavity pulse suppression might be required for longer lasers \cite{rosales12}. Polaritonic systems with very strong Kerr interactions are hence expected to be more robust.

\section{Conclusion}

In summary, we have presented the demonstration of a pulsed polariton laser operating from $\text{T}=150 \ K$ to room temperature, and relying on a mode-locking mechanism based on the strong polariton-polariton Kerr non-linearity. This last non-linearity is strong enough to compensate for the large group velocity dispersion associated to the lower polariton branch in the strong exciton-photon coupling regime. The chosen cavity geometry, with a GaN-based active layer, is highly similar to standard ridge semiconductor edge-emitting lasers, but our investigation as a function of the gain length highlights the strong difference with the saturable absorber approach to mode-locking. This design is compatible with standard electrical schemes and with further integration into nitride-based photonic circuits targeting the NUV-VIS spectral range. The spectral signatures of mode-locking are well reproduced through simulations solving the Gross-Pitaevskii equation for the coupled exciton and photon fields, leading to $100~fs~-$ solitons. The detailed modelling at $\text{T}=150 \ K$ leads to an estimated $5 ~ fJ$ soliton energy per pulse, well-suited to applications requiring low-energy laser pulses. 
Overall, the compact micrometric footprint, the laser repetition frequency beyond $300 ~ GHz$ and the room-temperature operation confer these polariton-based pulsed laser promising figures of merit for the development of UV integrated photonics sources. 

\section*{Funding}
Agence Nationale de la Recherche (ANR-11-LABX-0014, ANR-16-IDEX-0001, ANR-21-CE24-0019-01); Horizon 2020 Framework Programme (964770, FET 964770); Région Occitanie Pyrénées-Méditerranée (ALDOCT-001065).

\section*{Acknowledgments}
The authors acknowledge fundings from the French National Research Agency and the Region Occitanie, and the support of the  European Union’s Horizon 2020 program, through a FET Open research and  innovation action. C2N is a member of RENATECH-CNRS, the French national  network of large micro-nanofacilities.

\section*{Data availability}
Data underlying the results presented in this paper are available in Ref.~ \cite{OpenDataSet}

\appendix
\section{Cavity imaging}\label{app:Cavity imaging}
Microphotoluminescence imaging ($\mu$PL) is employed for optical investigations. The cavity undergoes excitation at its center using a pulsed laser ($355~nm$, $7~kHz$ frequency, $4~ns$ pulses), resonantly activating the exciton reservoir through a line-shaped spot profile. The adjustable length, $L_\text{pump}$, enables the partial or complete excitation of the cavity, thereby regulating the size of the exciton reservoir. A single microscope objective was employed for the injection and collection of light, facilitated by the use of a beamsplitter. The spectrometer, equipped with a diffractive grating, is connected to a charge-coupled device (CCD) to reconstruct the real-space image. This allows to make a clear distinction between the emission associated to the excitonic reservoir (located at the cavity center) and the emission of guided polaritons (positioned at the Distributed Bragg Reflectors). This latter is marked by a series of sharp peaks corresponding to the Fabry-Perot (FP) modes of the cavity.

\section{Strong coupling regime and group velocity dispersion} \label{app:Strong coupling regime and group velocity dispersion}
The strong coupling regime is established from the energy dispersion of the modes, measured either through gratings \cite{Brimont_Strong_2020} or from the free spectral range (FSR) separating the cavity modes. The Taylor expansion of the dispersion, as defined by~\cite{Agrawal_Nonlinear_2001}, writes:
\begin{equation}
\beta(\omega)=n(\omega) \frac{\omega}{c}=\beta_0+\frac{1}{v_\text{g}} (\omega-\omega_0)+\frac{1}{2}\beta_2(\omega-\omega_0)^2+... .
\end{equation}
where $\beta$ is the propagation constant in the waveguide, $n(\omega)$ is the refractive index, $v_\text{g}$ is the group velocity and $\beta_2$ is the group velocity dispersion parameter (GVD), which characterizes the pulse broadening phenomenon caused by the dispersion in the group velocity. 

The FSR is directly proportional to the first derivative of the polariton energy in the lower {TE0} branch, {\it i.e.} to  $v_\text{g}:$
\begin{equation}
  \text{FSR} = \frac{\pi}{L_\text{cav}} \ \frac{\partial E_\text{LPB}}{\partial \beta} = \frac{h}{2 \ L_\text{cav}} v_\text{g}, 
  \label{FSR}
\end{equation}

The achievement of the strong coupling regime can be assessed by comparing the measured cavity FSR to the theoretical FSR with and without exciton resonances, which are extracted from the dispersions deduced on the Elliott-Tanguy model of the dielectric susceptibility~\cite{Brimont_Strong_2020} (see Fig.~\ref{fig:dispersion}(d)). {The other polariton branches (TE1, TM0) are not observed as additional series of modes in the spectra, probably because of larger losses leading to a poorer mode finesse.} In this model, the GaN exciton energy $E_\text{XA} = 3.475 \ eV$ is determined from the reflectivity spectra of the same active layer taken at 150$\ K$, whereas the core and cladding thicknesses are measured from scanning electron micrographs.

\section{Nonlinear refractive index and polariton-polariton interaction constant}\label{app:Nonlinear refractive index and polariton-polariton interaction constant}
The refractive index $n$ can be expressed as the sum of a linear term $n_0$ and a non-linear term, which is typically related to the non-linear refractive index coefficient $n_2$ and the electric field amplitude $F$~\cite{Agrawal_Nonlinear_2001,belanger_solitonlike_1997}: $n=n_0+n_2 |F|^2 / A_\text{eff},$
where $A_\text{eff} \approx 0.15~\mu m^2$ is the transverse effective area of the guided modes (the product of the active layer thickness $150 \ nm$ and the ridge width $1 \ \mu m$). $n_2$ is indeed a measure of the strength of non-linearity in the material.\\
In the case of a Kerr medium, $n_2$ is related to the real part of third-order non-linear susceptibility $\chi^{(3)}$ \cite{Agrawal_Nonlinear_2001}: $n_2=3 \ \mathrm{Re}(\chi^{(3)}) \ / (8 n_0).$
It imposes a phase modulation of the propagating wave along a wire $n_2 (|F|^2 / A_\text{eff}) \beta_0 x$ where $\beta_0$ is the wave vector and $x$ the coordinate.\\
In the Gross Pitaevskii picture describing non-linearities in terms of interaction between particles, the interaction energy $\alpha N$ induces a phase modulation $\alpha N t$ where $t$ is the propagation time. By writing that position and propagation time are linked by the phase velocity $x=ct/n$ allows to express $n_2$ as:
{ \begin{equation}
n_2=-2n^2\alpha V/E_\text{LPB}^2c.
\end{equation}
}
$\alpha$ is the polariton-polariton interaction constant.  It can either be deduced from the FSR shift measured in Fig.~\ref{fig:Exp_nonlinearity}: 
\begin{equation}
\text{FSR}(N)-\text{FSR}(0) = \alpha N/2\sqrt{2}
\label{FSR-N}
\end{equation}
or from its theoretical expression 
\begin{equation}
\alpha = x_\text{p}^2 \alpha_\text{X,1D} / (T_0 v_\text{g})
\label{alpha}
\end{equation}
Here $x_\text{p}\approx 1/4$ is the exciton fraction of the polariton mode; $T_0 v_\text{g} \approx 4 \ \mu m$ is the soliton healing length along the ridge, where the pulse duration $T_0$ is deduced from spectra (see Appendix~\ref{app:Secant hyperbolic pulse propagation in a dispersive nonlinear material}). The 1D exciton-exciton interaction constant relates to its 3D counterpart: $\alpha_\text{X,1D} = \alpha_\text{X,3D} / A_\text{eff}$ \cite{haug1977theory} with $\alpha_\text{X,3D}=6.6 \pi E_\text{b} a_\text{B}^3$ ($E_\text{b}=28~meV$ is the exciton binding energy and $a_B=3.5~nm$ the exciton Bohr radius).

\section{Secant hyperbolic pulse propagation in a dispersive nonlinear material}\label{app:Secant hyperbolic pulse propagation in a dispersive nonlinear material}
Let us here consider the propagation of a secant hyperbolic pulse, with the envelope of the electric field amplitude given by: 
\begin{equation}
F(t)\propto \mathrm{sech} \left (  \frac{t}{T_0} \right ).
\end{equation}
$T_0$ measures the pulse duration, such that the full width at half maximum (FWHM) of the pulse can be expressed as follows~\cite{Agrawal_Nonlinear_2001}:
\begin{equation}
T_\text{FWHM}=2\ ln(1+\sqrt{2})\ T_0\approx 1.763\ T_0 .
\label{FWHM}
\end{equation}
The corresponding spectral lineshape is
\begin{equation}
I(E)\propto \left (\mathrm{sech} \left (  \frac{\pi \ T_0 \ E}{2 \hbar} \right )\right)^2 .
\label{eq:sech}
\end{equation}
Two characteristic length can be deduced, which characterize the respective impacts of dispersion and non-linearity:
\begin{itemize}
    \item The dispersion length $L_\text{D}$, a characteristic distance over which the effects of dispersion become significant~\cite{Agrawal_Nonlinear_2001}:
\begin{equation}
L_\text{D} = \frac{T_0^2}{| \beta_2 |} ,
\end{equation}
    \item The non-linear length $L_\text{NL}$ is defined for a given peak power $P_0$ as~\cite{belanger_solitonlike_1997}: 
\begin{equation}
L_\text{NL} = \frac{n_0\ A_\text{eff}}{\beta_0 \ | n_2 | \ P_0}.
\end{equation}
\end{itemize}

\section{\textit{CW} lasing at T=70~K}\label{app:CW lasing at T=70K}
The competition between lasing modes at $\text{T} = 70 \ K$ depends on the pump length. It is observed that the contrast between the first lasing mode and the next ones is stronger if the pump length is shorter. Figure~\ref{fig:Monomode_small_spot} presents the spectra of the same cavity as in Figure~\ref{fig:Single_Multimode}, but pumped over a smaller pump length $L_\text{pump} = 6.5 \ \mu m$. The laser emission remains monomode up to $P = 1.4 \times P_\text{th}$.

\begin{figure}[ht]
\centering{\includegraphics[width=7cm]{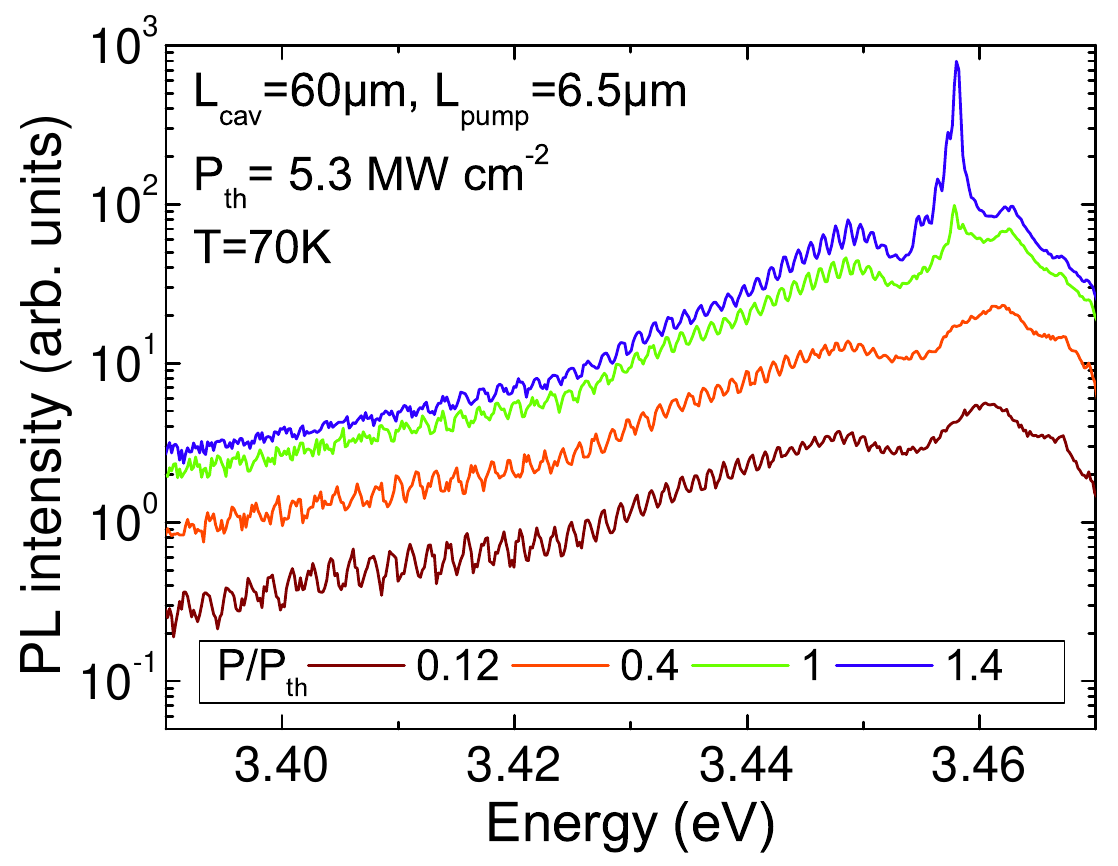}} 
  \caption{ Laser operation of a $60 \ \mu m-$long cavity and a short pump length $L_\text{pump} = 6.5 \ \mu m $: Power-dependent emission spectra across the lasing threshold, at $\text{T} = 70 \ K$ {(Sample~ A)}.
  }
  \label{fig:Monomode_small_spot}
\end{figure}

\section{Multimode lasing at T=150~K} \label{app:Multimode lasing at T=150K}

\subsection{Evolution of the FSR} 
The Figure~\ref{fig:FSR}(a) presents the evolution of the FSR over a subset of spectra vs the pump power $P$. The full series of data is presented in the Figure~\ref{fig:FSR_waterfall}. Each dataset is shifted by $+0.3 \ meV$ in a waterfall representation in order to emphasize the transition from a monotonic decrease of the FSR vs energy below threshold, to a more complex dependence with a flattened range (between the gray lines) above threshold.

\begin{figure}[htbp]
\centering{\includegraphics[width=7cm]{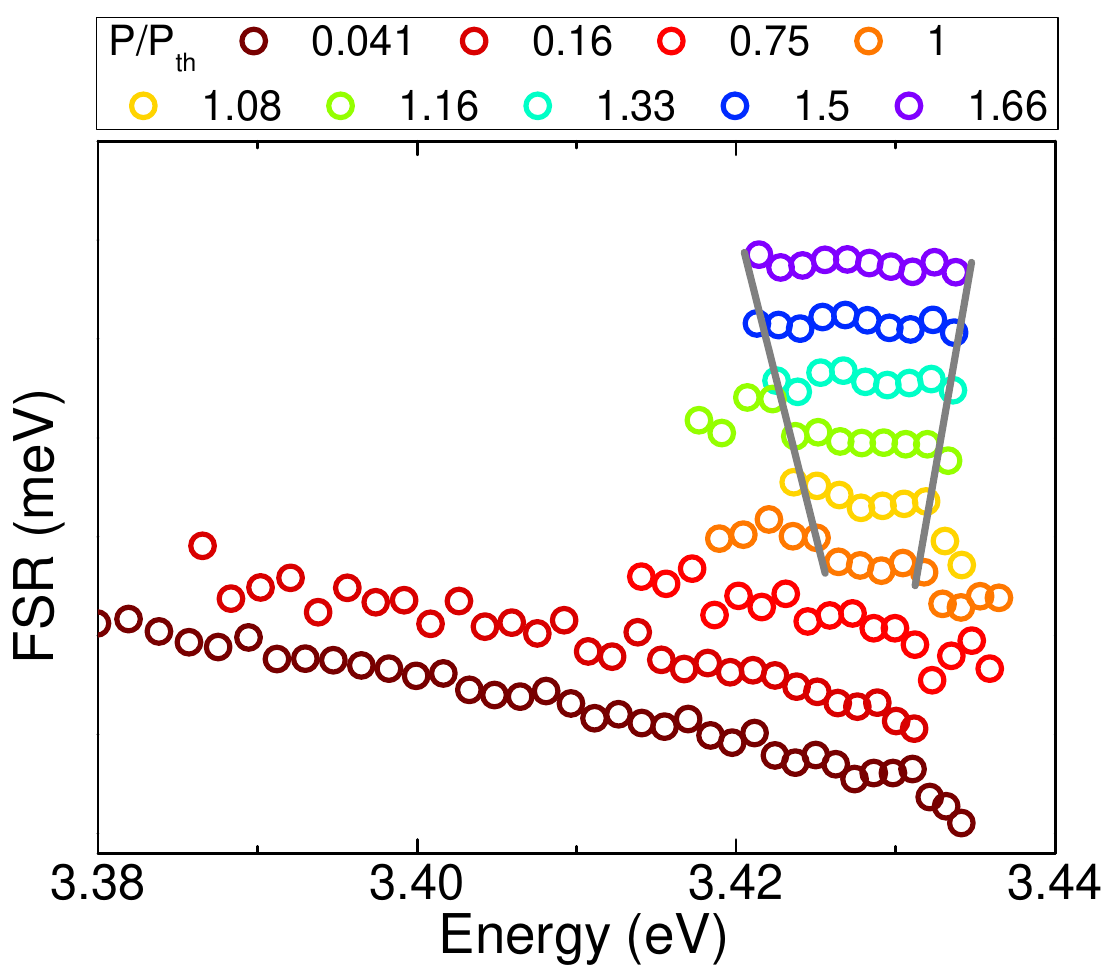}} 
  \caption{ 
  Free Spectral Range (FSR) of the modes {(Sample A)} across the lasing threshold, at $T=150 \ K.$ The gray lines are guides for the eye indicating the lasing modes with a constant FSR.}
  \label{fig:FSR_waterfall}
\end{figure}

\subsection{Quantifying the weakening of the exciton-photon interaction below threshold}

\begin{figure}[htbp]
\centering{\includegraphics[width=7.5cm]{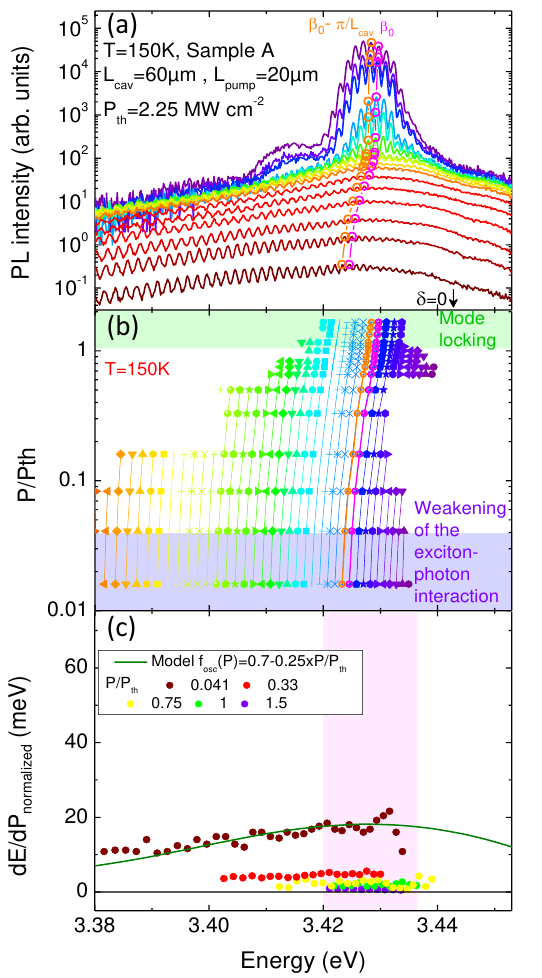}} 
  \caption{{(a) Power-dependent emission spectra across the lasing threshold, at $\text{T}=150 \ K$ (same as Fig.~\ref{fig:Single_Multimode}(b)), and peak-picking of the 2~modes associated to the quantized wavevectors $\beta_0$ and $\beta_0 - \pi/L_\text{cav}$} (b) Energy of each mode of the polariton laser as a function of the pump power; the lasing modes are indicated by the vertical pink stripe; the blueshift is dominated by the weakening of the oscillator strength in the bottom blue-shaded rectangle, and by the mode-locking dynamics in the top green-shaded area. (c) Derivative of the blueshift vs pump power, with the fit of the data at $P = 0.041\times P_\text{th}$ with the model accounting for the reduction of the exciton oscillator strength.
  }
  \label{fig:Weakening_fosc}
\end{figure}

\begin{figure*}
\centering{\includegraphics[width=12cm]{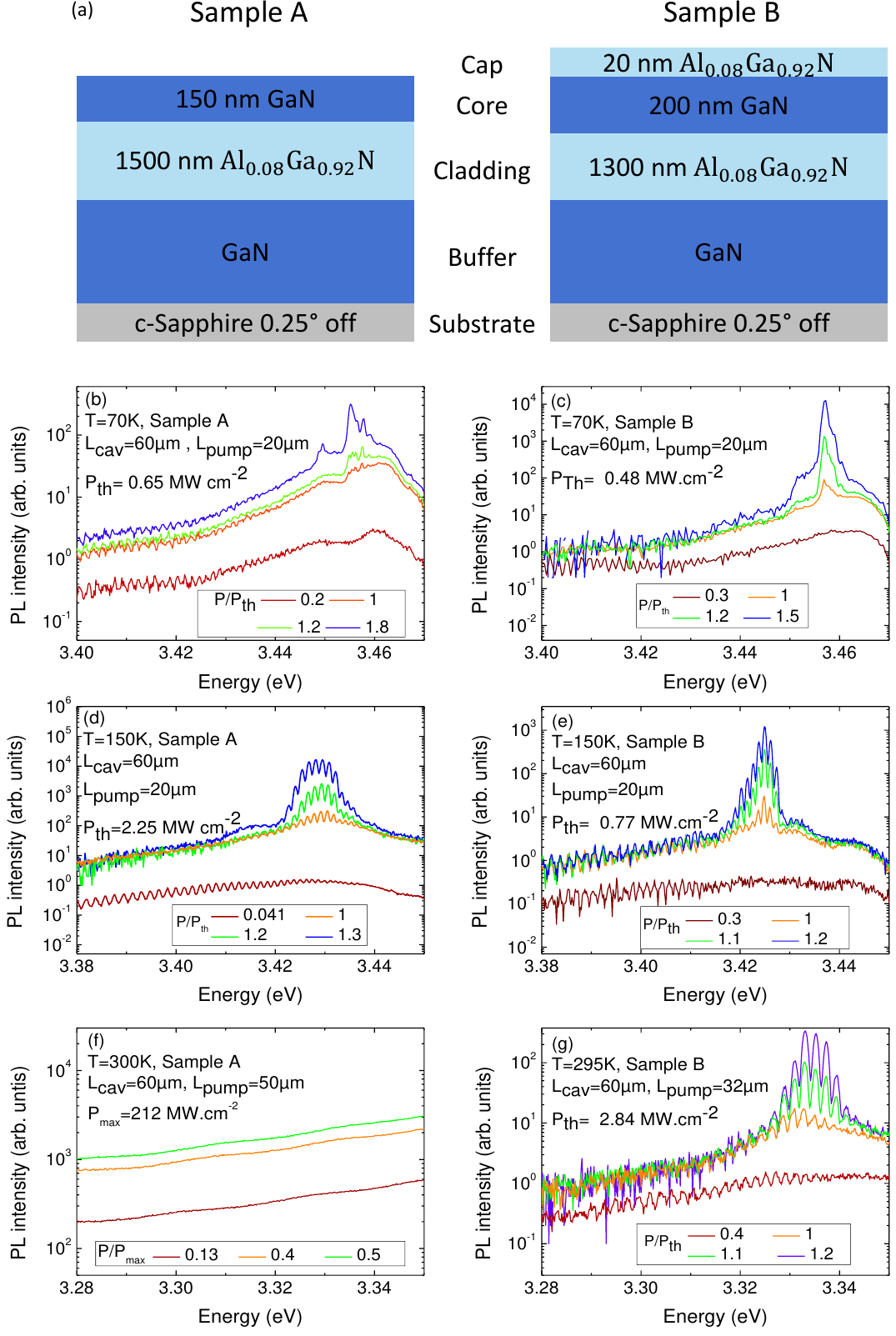}} 
  \caption{(a) Heterostructure design of samples A and B. {Evolution of the laser emission from $\text{T}= 70\ K$ to room-temperature for sample A (b,d,f) and sample B (c,e,g)}}
  \label{fig:samples}
\end{figure*}

Figure~\ref{fig:Weakening_fosc}(b) presents the measured shift of each individual polariton mode across the lasing threshold. Here we focus our analysis on the low power regime, for $P/P_\text{th} < 0.1$ typically, and consider the assumption of a progressive screening of the exciton oscillator strength due to the increased density of excitons in the reservoir, leading to an overall blueshift of the LPB dispersion. This effect is well documented for microcavity polaritons \cite{bajoni_photon_2007,Vladimirova_Polarization_2009,vladimirova_polariton-polariton_2010,bajoni_polariton_2012} and more recently for waveguide polaritons \cite{jamadi_edge-emitting_2018,Walker_Spatiotemporal_2019,Mallet-Dida_low_2022,Kreyder_Lasing_2022}. In such a case, this blueshift is not constant along the LPB branch: it reaches its maximum near zero exciton-photon detuning. Figure~\ref{fig:Weakening_fosc} presents (b)~the energy of each Fabry-Perot mode (extracted from the spectra in Fig.~\ref{fig:Single_Multimode}(a)) and (c)~its first derivative with respect to the normalized pump power $P / P_\text{th}$. At the lowest investigated pump power, the blueshift strongly depends on the exciton-photon detuning. Let us compare this feature to a first order model of the weakening of the oscillator strength: we already know from our previous works that the oscillator strength of the GaN excitons is very close to its nominal value in the case of a planar waveguide (correction factor $\times 0.95)$ \cite{Brimont_Strong_2020} whereas it is reduced by $30\% $ in the case of the presently investigated etched GaN waveguides (correction factor $\times 0.7$, \cite{Souissi_Ridge_2022} and Fig.~\ref{fig:FSR}). If we assume a linear decrease of the correction factor of the oscillator strength with the density of the exciton reservoir, i.e. with the pump power density, $f_\text{osc} (P) = f_\text{osc}^0 - \alpha P / P_\text{th},$ then we can calculate the new LPB dispersion, the corresponding blueshift for the Fabry-Perot modes and its derivative vs pump power. We reach a quantitative agreement with the experimental values (brown points) at the lowest pump power $P \le 0.041\times P_\text{th}$ for $\alpha = 0.25,$ i.e. a correction factor $f_\text{osc} (P) = 0.7 - 0.25 \times P / P_\text{th}$ (green plain line). When the pump power is further increased, we observe that the first derivative strongly deviates from this model, with a flat part close to zero around the lasing energy, and a steep change towards negative (resp. positive) values for the modes at energies below (resp. above) the lasing spectral range. We can therefore conclude that a weakening of the oscillator strength contributes to the blueshift of the modes. It fully explains the blueshift measured at the lowest investigated pump power well below threshold (blue shaded pump range in Fig.~\ref{fig:Single_Multimode}(a)); but it cannot explain the flattening of the modes rapidly taking place at threshold and attributed to mode-locking (green-shaded pump range).

\section{Multimode lasing at room temperature}  \label{app:Multimode lasing at room temperature}
The room temperature operation of the polariton laser requires an improvement of the initial epilayer design of sample~A: the sample~B embeds a thin cap layer improving the exciton confinement, and a slightly thicker core GaN layer. It is expected to reduce the surface recombination rate as the exciton diffusion increases at high temperature. The detailed heterostructures of samples~A and B are presented in the Figure~\ref{fig:samples}(a).

The laser dynamics of sample~B is similar to that of sample~A, which is presented and analyzed in detail in the manuscript, {at $\text{T}= 70\ K$ and $\text{T}= 150\ K$ as shown in the Figure~\ref{fig:samples}(b,c,d,e). However, at room-temperature the polariton modes are not observed in the spectra at low or high excitation power in sample~A (up to fifty times the threshold of sample~B) (Fig~\ref{fig:samples}(f));}, while the polariton modes and mode-locking are observed in sample~B (Fig~\ref{fig:samples}(g)). The laser dynamics of the sample~B is assessed both by the secant hyperbolic lineshape beyond threshold and the flattening of the FSR (Fig~\ref{fig:FSR_295K}(a) as well as the linearization of the polariton dispersion (Fig~\ref{fig:FSR_295K}b).

\begin{figure}[htbp]
\centering{\includegraphics[width=8cm]{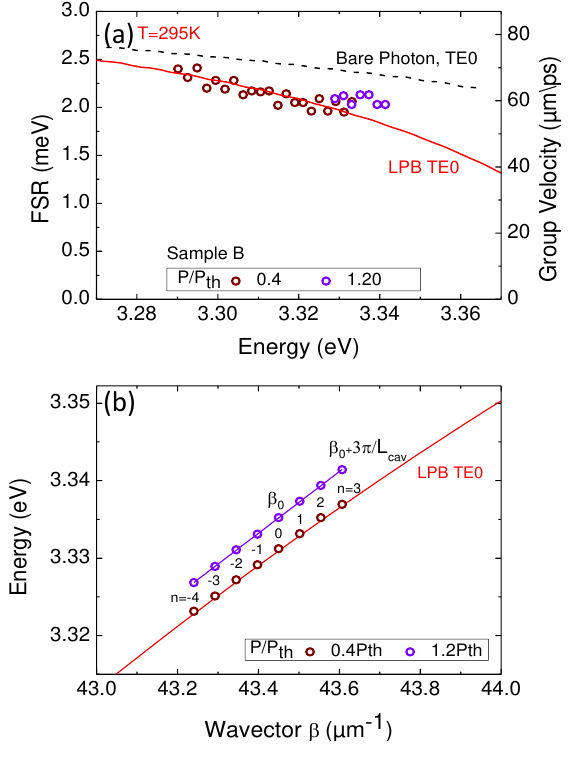}} 
  \caption{Experimental Free Spectral Range (circles) and group velocity $v_\text{g}$ of the modes across the lasing threshold, at $\text{T}= 295\ K$. The solid and dashed lines indicate the FSR deduced from the polariton dispersion and the bare photon mode calculated for the waveguide design of sample~B, respectively. (b)~Measured (circles) and modelled (lines) polariton dispersions below and above threshold. { The longitudinal modes are numbered relatively to the central mode $n=0$, and the quantized wavevectors $\beta_\text{n}$ for a 60$\ \mu m$ long cavity are indicated as circles.}}
  \label{fig:FSR_295K}
\end{figure}

\section{Numerical simulations}\label{app:Numerical simulations}

We use the 3rd order Adams-Bashforth method \cite{Bashforth1883} for numerical integration of the coupled 1D Gross-Pitaevskii equations~\eqref{GPE1},~\eqref{GPE2} for excitons and photons. The kinetic energy operator $\hat{T}$ is calculated using the Fourier transform implemented with parallel computing on the Central Processing Unit. The dispersion relations for excitons and photons are taken as in Fig.~\ref{fig:dispersion} of the main text. The gain operator 
$\hat{\Gamma}$ is evaluated using a similar procedure.

The high wave vectors involved in the experiment $k\approx 45\times 10^6~m^{-1}$, as compared with planar microcavities, where the typical values are of the order of $10^6~m^{-1}$, require the use of a relatively small spatial step size and therefore a small time step. We used a spatial step of $31.25~nm$ and a time step of $10^{-16}~s$.

\section{Dependence on the pump length}\label{app:Dependence on the pump length}
At $\text{T}= 150\ K$, the emission spectra of the 60-$\mu$m-long cavity, using a 50-$\mu$m-long pump spot are shown in Figure~\ref{fig:dependence on the pump length}(a) for a series of pumping powers. Far below threshold, the spectra are characterized by the presence of Fabry-Perot modes observed across a large energy range. Beyond threshold, a multimode emission resembling that in Figure~~\ref{fig:Single_Multimode}(b) is observed. The spectra above threshold can be effectively fitted using a secant hyperbolic squared function, as shown in Figure~\ref{fig:dependence on the pump length}(b). It is also important to note that the corresponding pulse duration, $T_0$, remains unchanged regardless of the spot length (compare Fig.~\ref{fig:dependence on the pump length}(b) and Fig.~\ref{fig:Single_Multimode}(b) in the main text).

\begin{figure}[htbp]
\centering{\includegraphics[width=8cm]{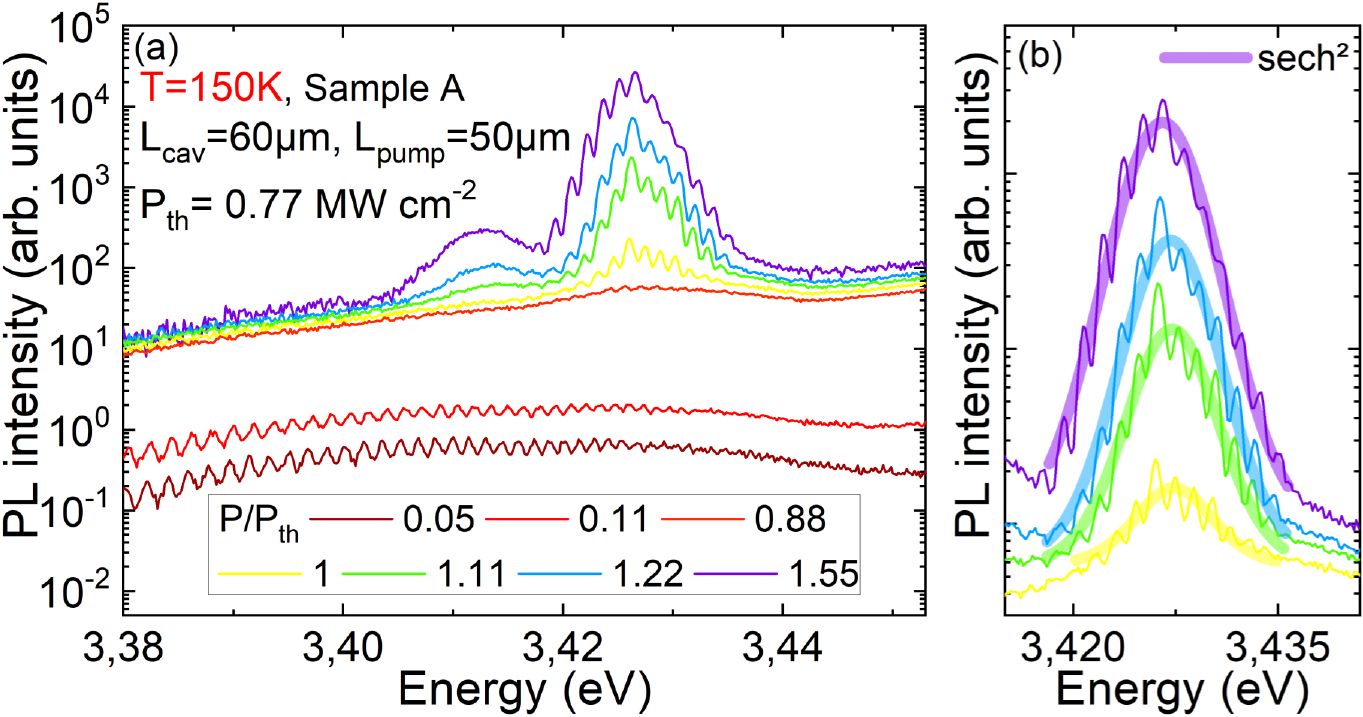}} 
  \caption{(a)~Emission at $\text{T}= 150\ K$ versus pumping power. The $60-\mu m$ length cavity is excited by a $50-\mu m$ sized pump. (b)~Secant hyperbolic squared fit of the envelope of the emission spectrum above threshold.}
  \label{fig:dependence on the pump length}
\end{figure}

%\bibliographystyle{./apsrev4-1} 
%\bibliography{Biblio-HSouissi}
\bibliography{Extraction_Biblio,biblio}

\end{document}